%% file: nu_2018_0vbb.tex
\documentclass[nofootinbib,superscriptaddress,twocolumn,showpacs,preprintnumbers,amsmath,amssymb,bibnotes,altaffilletter]{revtex4-1}
\usepackage{graphicx}
\usepackage{color}

\def\nuc#1#2{${}^{#1}$#2}
\def\mee{$\langle m_{\beta\beta} \rangle$}
\def\nonubb{$\beta\beta(0\nu)$}

\def\qval{$Q_{\beta\beta}$}
\def\MJ{{\sc Majorana}}
\def\DEM{{\sc Demonstrator}}
\def\MJD{{\sc Majorana Demonstrator}}

\begin{document}
\title{A Search for Neutrinoless Double-Beta Decay in \nuc{76}{Ge} with 26~kg-yr of Exposure from the \MJD}

\input{mjd_authors}

\begin{abstract}
The \MJ~Collaboration is operating an array of high purity Ge detectors to search for the neutrinoless double-beta decay of $^{76}$Ge.  The \MJD~consists of 44.1~kg of Ge detectors (29.7~kg enriched to 88\% in $^{76}$Ge) split between two modules constructed from ultra-clean materials.  Both modules are contained in a low-background shield at the Sanford Underground Research Facility in Lead, South Dakota.  We present updated results on the search for neutrinoless double-beta decay in $^{76}$Ge with $26.0\pm0.5$~kg-yr of enriched exposure.  With the \DEM's unprecedented energy resolution of 2.53~keV FWHM at \qval, we observe one event in the region of interest with 0.65 events expected from the estimated background, resulting in a lower limit on the $^{76}$Ge neutrinoless double-beta decay half-life of $2.7\times10^{25}$~yr (90\% CL) with a median sensitivity of $4.8\times10^{25}$~yr (90\% CL).  Depending on the matrix elements used, a 90\% CL upper limit on the effective Majorana neutrino mass in the range of 200-433~meV is obtained.  The measured background in the low-background configurations is $11.9\pm2.0$~counts/(FWHM~t~yr).
\end{abstract}


\maketitle

\section{Introduction}
\label{sec:intro}

A large, international experimental program is underway in search of neutrinoless double-beta decay (\nonubb); see~\cite{avi08, bar11, Rode11, Elliott2012, Vergados2012, Cremonesi2013, Schwingenheuer2013, Elliott2015, Henning2016, DBDReview2015, DBDReview2018} for example.  While the process of double-beta decay with the emission of two neutrinos has been experimentally observed in several nuclei, \nonubb~remains unobserved with half-life limits exceeding $10^{26}$~yr for some isotopes. Observation of \nonubb~would establish physics beyond the Standard Model by demonstrating lepton number violation and the Majorana nature of the neutrino~\cite{sch82}.  Furthermore, measurement of the \nonubb~decay rate, which depends on the effective Majorana mass (\mee), would help constrain the absolute neutrino mass scale in the light neutrino exchange model~\cite{Vergados2016}.

The next-generation of \nonubb~experiments aim to probe \mee~down to the level of $\sim15$~meV, the minimum allowable mass assuming the inverted neutrino mass ordering scenario~\cite{Ahm04}.  When combined with results from other neutrino experiments, \nonubb~experiments will begin to shed light on the nature of the neutrino, even in the case of non-observation.  For normal ordering, next-generation experiments will have \nonubb~discovery sensitivity in an interesting region of parameter space, and non-observation would improve existing limits by $\sim$1 order of magnitude~\cite{Agostini2017a}.  In order to probe \mee~at the level of $\sim$15~meV, next-generation experiments will need to deploy a $\sim$tonne of isotope in a nearly background-free experiment.

The \MJ~Collaboration is searching for \nonubb~decay in $^{76}$Ge using modular arrays of high-purity Ge (HPGe) detectors~\cite{Abgrall2014}.  The \MJD~is an array of 58 HPGe detectors built with the goal of demonstrating backgrounds low enough to justify construction of a tonne-scale Ge-based experiment.  $^{76}$Ge-enriched HPGe detectors are well-suited for \nonubb~searches due to the intrinsic purity of Ge, the well-understood Ge enrichment process, their excellent energy resolution, and their ability to perform pulse-shape-based particle identification.  The combined Ge mass of the \DEM~is 44.8~kg with 14.4~kg of natural Ge detectors and 29.7~kg of detectors enriched to 88.1$\pm$0.7\% in \nuc{76}{Ge}~\cite{Abgrall2018}.  The enriched detectors are P-type, point contact (PPC) detectors~\cite{luk89,Barbeau2007,Aguayo2011} with low capacitance and sub-keV energy thresholds, permitting a variety of low-energy physics studies~\cite{Abgrall2017a,MJD_LIPs}.  The ultra low-backgrounds achieved with the \DEM~also permit other searches for new exotic physics, for instance tri-nucleon decay~\cite{mjd_baryon_decay}.  Results from an initial search for \nonubb~in \nuc{76}{Ge} using an exposure of 9.95~kg-yr with the \DEM~are presented in~\cite{Aalseth2018}.  Adding to this initial exposure, here we present results from a total enriched Ge exposure of 26.0~kg-yr, including 11.8~kg-yr of newly unblinded data.

The \DEM's enriched detectors range in mass from 0.5 to 1.1~kg.  The processes used by \MJ~to convert the enriched Ge material into detectors achieved an unprecedented yield of 69.8\%~\cite{Abgrall2018}.  As described in Section~\ref{sec:ecal}, these detectors have achieved an energy resolution of $2.53\pm0.08$~keV at 2039~keV, the double-beta decay Q-value (\qval).  The \MJD~utilizes a number of ultra-low activity materials and methods to reduce environmental backgrounds~\cite{Abgrall2016}, including the use of a total of 1196~kg of underground electroformed copper (UGEFCu) to construct the detector support structures, the cryostats, and the inner-most 5~cm of shielding surrounding the cryostats.  Carefully selected, commercially available low-background materials were used for the cabling, cryostat seals, and wherever electrical and thermal insulation was required.  Low-background front-end electronics were developed for the \DEM~as described in~\cite{Barton2011,Guinn2015}.
  
The \DEM's detectors are split between two modules contained in a low-background shield.  A 5-cm-thick layer of commercially sourced C10100 copper surrounding the inner UGEFCu shield provides additional shielding.   The copper shielding is contained within 45~cm of high-purity lead shielding, separating the low-background environment from the higher background electronics, cryogenic and vacuum hardware, and laboratory environment.  The lead shield is enclosed within a radon exclusion volume that is purged continuously with liquid-nitrogen boil-off gas.  An active muon veto~\cite{Bugg2014} surrounds the radon exclusion volume, and is itself enclosed in 5~cm of borated polyethylene and 25~cm of polyethylene for neutron moderation.  The shielded volume and all data acquisition (DAQ) and control electronics are situated in a clean room in the Davis Campus of the Sanford Underground Research Facility (SURF)~\cite{Heise2015} in Lead, South Dakota, at the 4850-foot level (4300~m.w.e).  The parts tracking database used to monitor cosmogenic exposure and inventory is described in~\cite{Abgrall2015}.

Each module is equipped with a system to deploy a \nuc{228}{Th} line source into the shield for periodic ($\sim$weekly) calibrations.  When in use, the line source is contained within a helical tube surrounding the cryostat.  The calibration system is described in detail in~\cite{Abgrall2017}.

\section{Data and Event Selection}
\label{sec:analysis}

The data presented here are divided into seven data sets, referred to as DS0 through DS6 (detailed below and summarized later in Table~\ref{tbl:efficiency}).  Data set boundaries are defined by significant changes in the experimental configuration that occurred during construction and commissioning.  Minor changes to the experimental configuration or DAQ within a data set are distinguished by sub-ranges denoted by a letter following the data set number.

Data acquisition for DS0 began with module 1 on 26~July~2015 without the inner UGEFCu shield, without UGEFCu shielding along the vacuum penetration into the shield, and with higher activity Kalrez cryostat seals.  DS1 began when the final UGEFCu shielding was in place.  Module 1 continued to operate alone within the shield in DS2, when the digitizers were operated in a mode that pre-summed the region following the rising edge of the waveform to investigate potential improvements to alpha discrimination with the longer acquisition window (see Section~\ref{sec:dcr}).  Both modules were installed in the shield and operated simultaneously for DS3 (module 1) and DS4 (module 2), but the modules were controlled with independent DAQ systems without waveform pre-summing.  During DS5, both modules were operated using the same DAQ system.   DS5a had increased electronics noise during the integration of the DAQ systems, completion of the polyethylene shielding, and optimization of the grounding scheme.  DS5b and DS5c were acquired in the same hardware configuration, but the data blindness scheme was then imposed for DS5c.  In DS6, the waveform pre-summing was re-enabled.  The hardware configuration and blindness were otherwise unchanged from DS5c.  The DS6 data acquired up to 16 April 2018 is referred to as DS6a.

The \MJD~implements data blindness through a prescaling scheme in which 31~hours of open background data are acquired, followed by a 93~hour period of blind data.  All calibration data and data taken during maintenance and testing are exempt from the blindness scheme.  The data acquired in DS5c and DS6a were not included in the analysis presented in~\cite{Aalseth2018}.  Additionally, the blind data acquired during DS1 and DS2, which was not opened for the result in~\cite{Aalseth2018}, is analyzed here.  Table~\ref{tbl:efficiency} summarizes the starting date, enriched detector mass, and exposure for each data set.

The Ge detector signals are digitized with 14~bit, 100~MS/s digitizers~\cite{Anderson2009} originally developed for the GRETINA experiment~\cite{Vet00}.  The waveforms are recorded in a 20-$\mu$s acquisition window at the full sampling rate with the window divided evenly into the pre- and post-trigger regions.  In DS2 and DS6, each recorded sample in the region 4~$\mu$s after the rising edge is the pre-summed value of four subsequent 10~ns samples, increasing the total acquisition window to 38.2~$\mu$s.  Each detector has a high-gain and a low-gain signal amplification path that are digitized independently.  The trigger threshold for each channel is set independently based on its trigger rate, which depends both on the electronic noise and the initialization of the on-board trapezoidal filter.  Due to firmware limitations, in cases where the initialized value of the on-board filter used for triggering is negative (due to electronics noise or baseline recovery from interactions in the detectors at the time of initialization) a small dead time may be induced, which is incorporated into the live time calculation for each data set.  This results in a detector-dependent reduction in live time, typically $<0.1$\%, that is estimated using the fraction of periodic pulser signals triggering each channel.

In offline analysis, recorded waveforms are grouped into physics events using a 4-$\mu$s coincidence window.  Since \nonubb~events are confined to a single detector and are contained completely by the digitization window, events in which multiple detectors trigger are rejected.  Each waveform is then checked against a set of data quality metrics that eliminate non-physical waveforms and signals from periodic pulsers.  The acceptance of this `data cleaning' procedure for physics events is estimated to be $>99.9$\% for all data sets.  Events within 1~s of a trigger from the muon veto system are also rejected.  The data collected during liquid nitrogen fills, which occur every $\sim$36~hr, are discarded due to microphonic noise.  Each fill results in approximately 30~min of rejected data for the module being filled.

\section{Energy Estimation}
\label{sec:edet}

The energy of each event is estimated using standard Ge detector techniques that measure the calibrated amplitude of filtered, pole-zero corrected signals (see~\cite{Jordanov_Knoll} for example).  Then, finely tuned and calibrated corrections that account for ADC non-linearities and charge trapping along the drift path are incorporated to achieve the measured 2.53~keV resolution at the 2039~keV Q-value, for the exposure-weighted combination of the enriched detectors.

\begin{figure}[!htbp]
  \centering
  \includegraphics[width=\columnwidth]{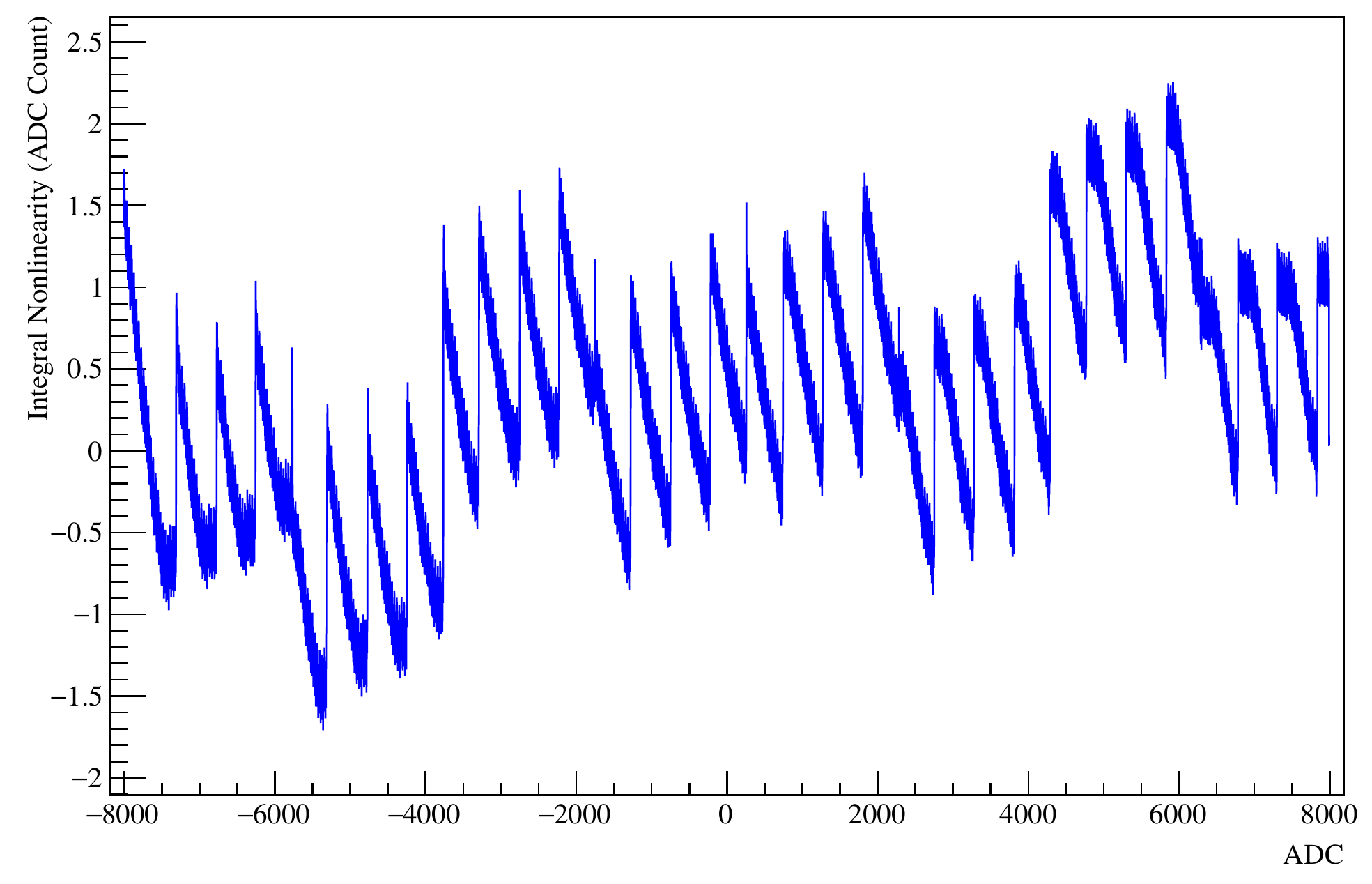}
  \caption{Color online. An example of the measured integral non-linearity for a digitizer channel in the \MJD.  The integral non-linearity deviation is less than $\sim2$ADC counts across the full range, but, without correction, this can result in as much as a 0.8~keV shift near the Q-value.\label{fig:nonlinearity}}
\end{figure}

The performance of multi-range ADC chips can depend both on the value and rate-of-change of the input voltage~\cite{adc_properties}. These effects have been measured for each of the GRETINA digitizer channels used in the \MJD~using external signals from waveform generators. Figure~\ref{fig:nonlinearity} shows the resulting integral non-linearity for a typical channel, which can amount to as much as a 0.8~keV shift in the estimated energy near \qval~for the high-gain readout.  Each waveform acquired is corrected for the measured integral non-linearity on the respective channel.  The correction reduces the energy uncertainty due to ADC effects to less than 0.1~keV based on comparison of the energy estimated using the high- and low-gain paths which have different ADC non-linearities.

Drift-path-dependent charge trapping in the crystal bulk also degrades the energy resolution~\cite{Martin2012}.  To account for charge trapping effects, the standard pole-zero correction is modified with an additional term that assumes exponential trapping of charges along the drift path.  The modified pole-zero time constant ($\tau$) is defined as
\begin{equation}
  \label{eq:chargetrapping}
  \frac{1}{\tau} = \frac{1}{\tau_{PZ}}-\frac{1}{\tau_{CT}},
\end{equation}
where $\tau_{PZ}$ is the pole-zero time constant due to the pre-amplifier (approximately 70~$\mu$s) and $\tau_{CT}$ is the correction for charge trapping effects.  For each detector, this modified pole-zero correction is optimized by minimizing the full-width at half maximum (FWHM) of the 2615~keV \nuc{208}{Tl} peak measured in calibration data.  The optimal value of $\tau_{CT}$ is typically near 233~$\mu$s resulting in an optimal value for $\tau$ near 100~$\mu$s.

\begin{figure}[!htbp]
  \centering
  \includegraphics[width=\columnwidth]{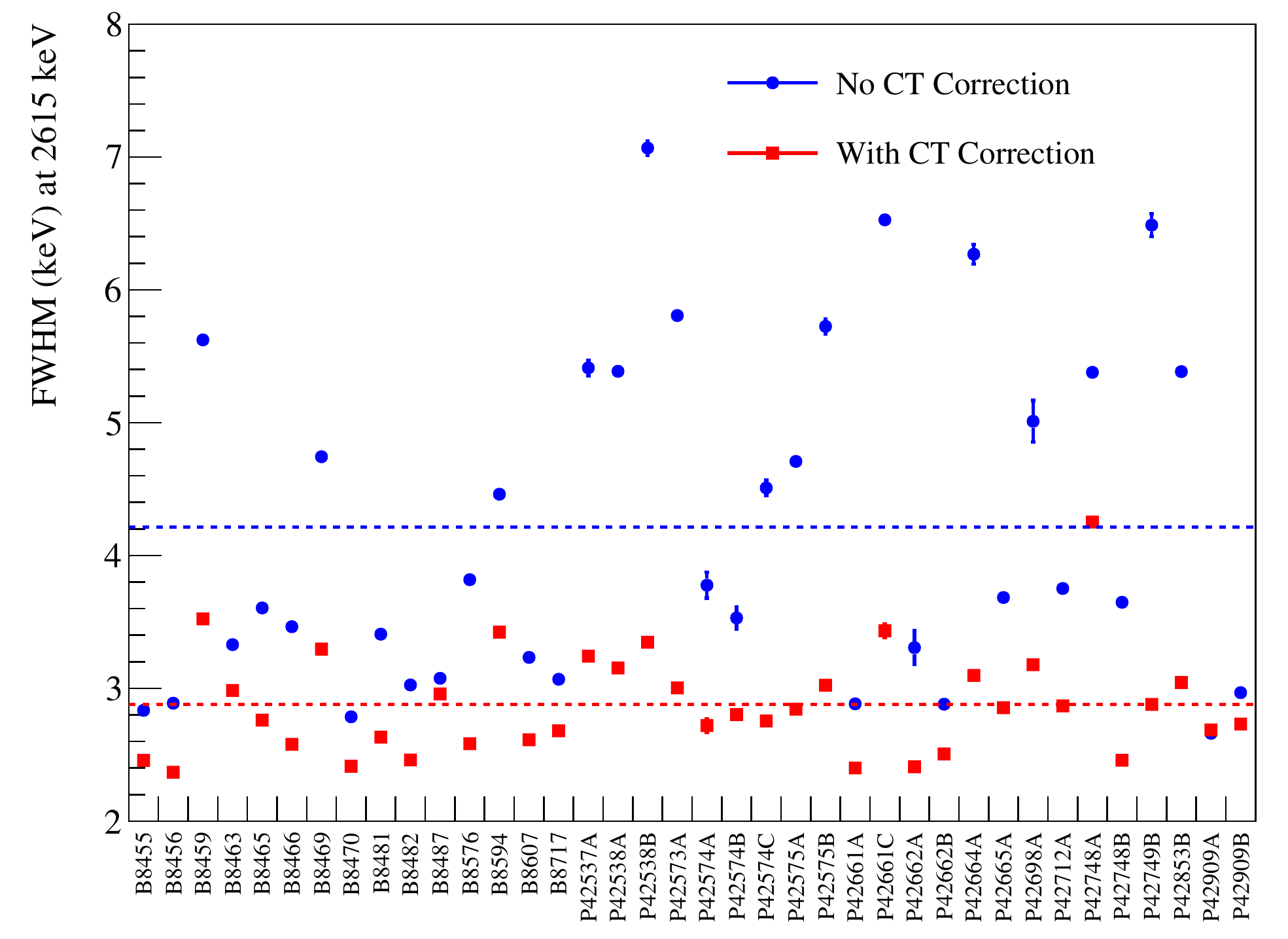}
  \caption{Color online. FWHM of the 2615~keV \nuc{208}{Tl} calibration peak from all operating detectors in DS6a before (blue points) and after (red points) the charge trapping correction.  The detector serial numbers are shown as the horizontal axis labels with natural detectors grouped on the left and enriched detectors on the right (serial numbers beginning with `B' and `P', respectively).  The horizontal lines indicate the mean resolution for all calibrated detectors, including natural detectors.\label{fig:chargetrapping}}
\end{figure}

The start time, $t_0$, of every waveform is estimated by applying a fast trapezoidal filter (1.0-$\mu$s rise time and 1.5-$\mu$s flat-top time) and determining the threshold crossing time.  The modified pole-zero correction and a slower trapezoidal filter (4.0~$\mu$s rise time and 2.5~$\mu$s flat top time) are then applied to the original waveform.  The value of the ADC-non-linearity-corrected, trapezoidal-filtered waveform at a fixed-time 0.5~$\mu$s from the end of the flat top, relative to $t_0$, is then used to estimate the energy of the event at a time beyond the region sensitive to charge trapping effects.  The use of the modified pole-zero correction results in a 1.4~keV improvement in energy resolution averaged over all operating detectors, as shown in Fig.~\ref{fig:chargetrapping} for the 2615~keV \nuc{208}{Tl} calibration peak.

\section{Energy Calibration}
\label{sec:ecal}

Periodic energy calibrations are used to provide an initial linear energy scale calibration for each channel based on the intrinsic resolution from electronics noise at zero energy and the position of the 2615~keV \nuc{208}{Tl} gamma line.  This provides an initial correction for any small variations over time in the electronics noise or energy scale.  In the second step of the calibration procedure, the statistics of the combined spectrum of all calibrations in each data set are sufficient to reliably perform a simultaneous fit to eight peaks in the calibration spectrum for a more finely-tuned energy calibration.  The full-energy gamma peaks used are at 239~keV (\nuc{212}{Pb}), 241~keV (\nuc{224}{Ra}), 277~keV (\nuc{208}{Tl}), 300~keV (\nuc{212}{Pb}), 583~keV (\nuc{208}{Tl}), 727~keV (\nuc{212}{Bi}), 861~keV (\nuc{208}{Tl}), and 2615~keV (\nuc{208}{Tl}).  Peaks due to single- and double-escape are excluded due to potential differences in peak shape.  Peaks of relatively low amplitude or in close proximity to other features in the calibration spectrum are also not used in the simultaneous fit.

\begin{figure}[!htbp]
  \centering
    \includegraphics[width=\columnwidth]{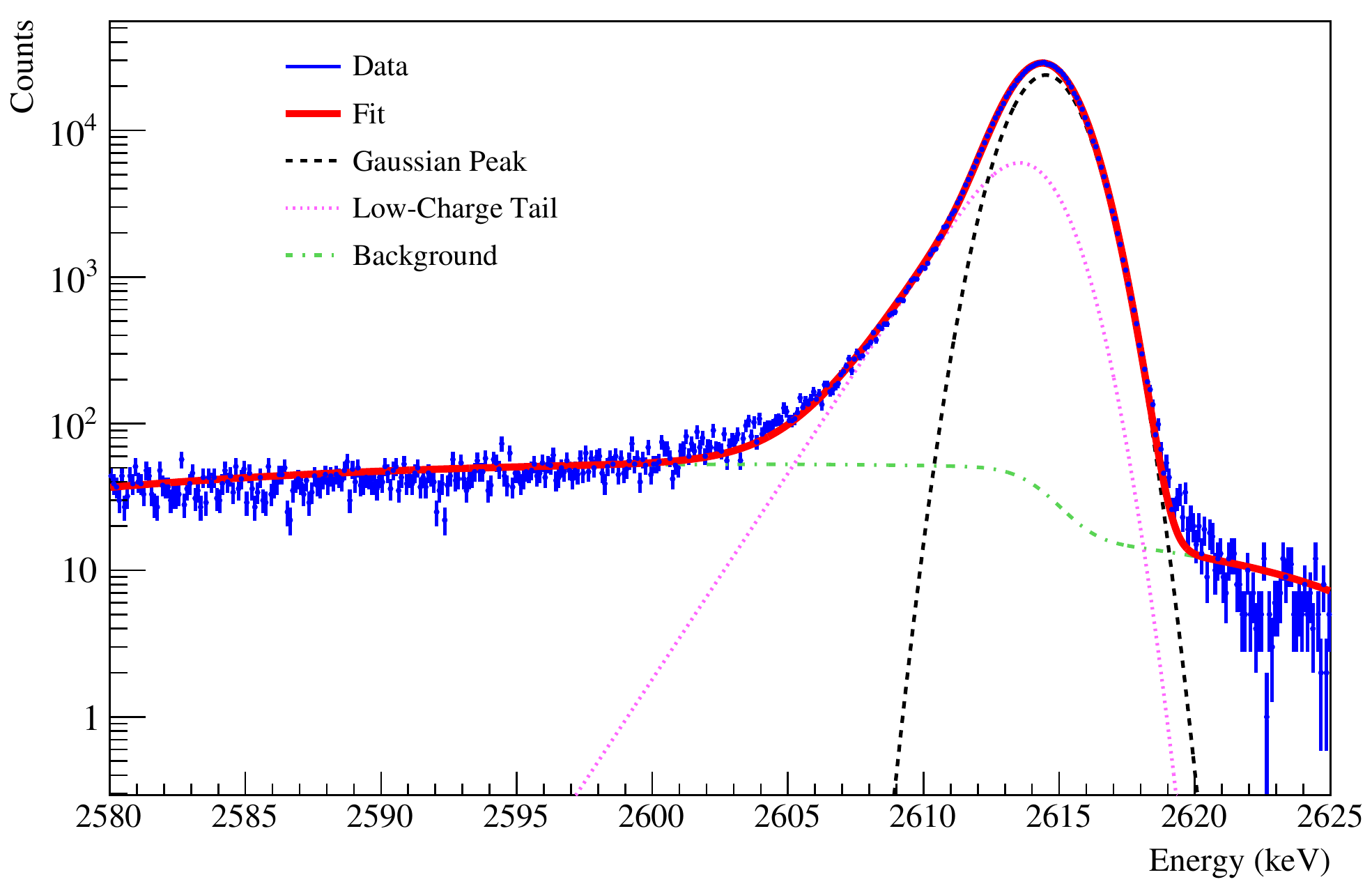}
    \caption{Color online. The 2615~keV peak from \nuc{208}{Tl} in calibration data with all detectors combined is shown in the blue points with statistical error bars.  The fitted background is shown in green, and the components of Eqn.~\ref{eq:peakshape} are shown in magenta and black.  The solid red curve shows the best fit to the sum of the background and peak shape.  The slight disagreement in the shape above the peak is due to inefficiencies in the rejection of pileup during calibration runs.\label{fig:peakfit}}
\end{figure}

The peak shape is modeled as the sum of a full-energy Gaussian component and an exponentially modified Gaussian tail to approximate the peak shape distortion due to  incomplete charge collection.  The response function ($R$) as a function of energy ($E$) for a mono-energetic line at energy $\mu$ is given by
\begin{equation}
  \label{eq:peakshape}
  \begin{split}
    R(E) = & \frac{1-f}{\sqrt{2\pi\sigma^2}}e^{-\frac{(E-\mu)^2}{2\sigma^2}} \\
               & + \frac{f}{2\gamma}e^{\left(\frac{\sigma^2}{2\gamma^2}+\frac{E-\mu}{\gamma}\right)}\text{erfc}\left(\frac{\sigma}{\sqrt{2}\gamma}+\frac{E-\mu}{\sqrt{2}\sigma}\right),
    \end{split}
\end{equation}
where $\sigma$ represents the smearing due primarily due to electronics noise and charge collection statistics, $\gamma$ is the decay constant of the low-energy tail, and $f$ is the fraction of the peak shape contained in the low-energy tail.  For each peak in the calibration data, the background in the vicinity of the peak is modeled by the sum of a complimentary error function shifted to the peak energy ($\mu$) with an underlying continuum component approximated with a quadratic polynomial.  The complimentary error function accounts for incident-particle-specific effects, like forward scattering, that are not related to the detector response.  The simultaneous fit to the peaks in the calibration spectrum is performed using Hybrid Monte Carlo~\cite{hybrid_mc}, a gradient-based Markov Chain Monte Carlo technique.  Figure~\ref{fig:peakfit} shows an example fit to the 2615~keV \nuc{208}{Tl} peak with all data sets and detectors combined.  The FWHM of the best-fit peak shape is 2.95~keV, and the value of $f$ in Eqn.~\ref{eq:peakshape}, the fraction of the peak in the low-energy tail, takes the value of 0.26.

\begin{figure*}[!htbp]
  \centering
  \includegraphics[width=0.8\textwidth]{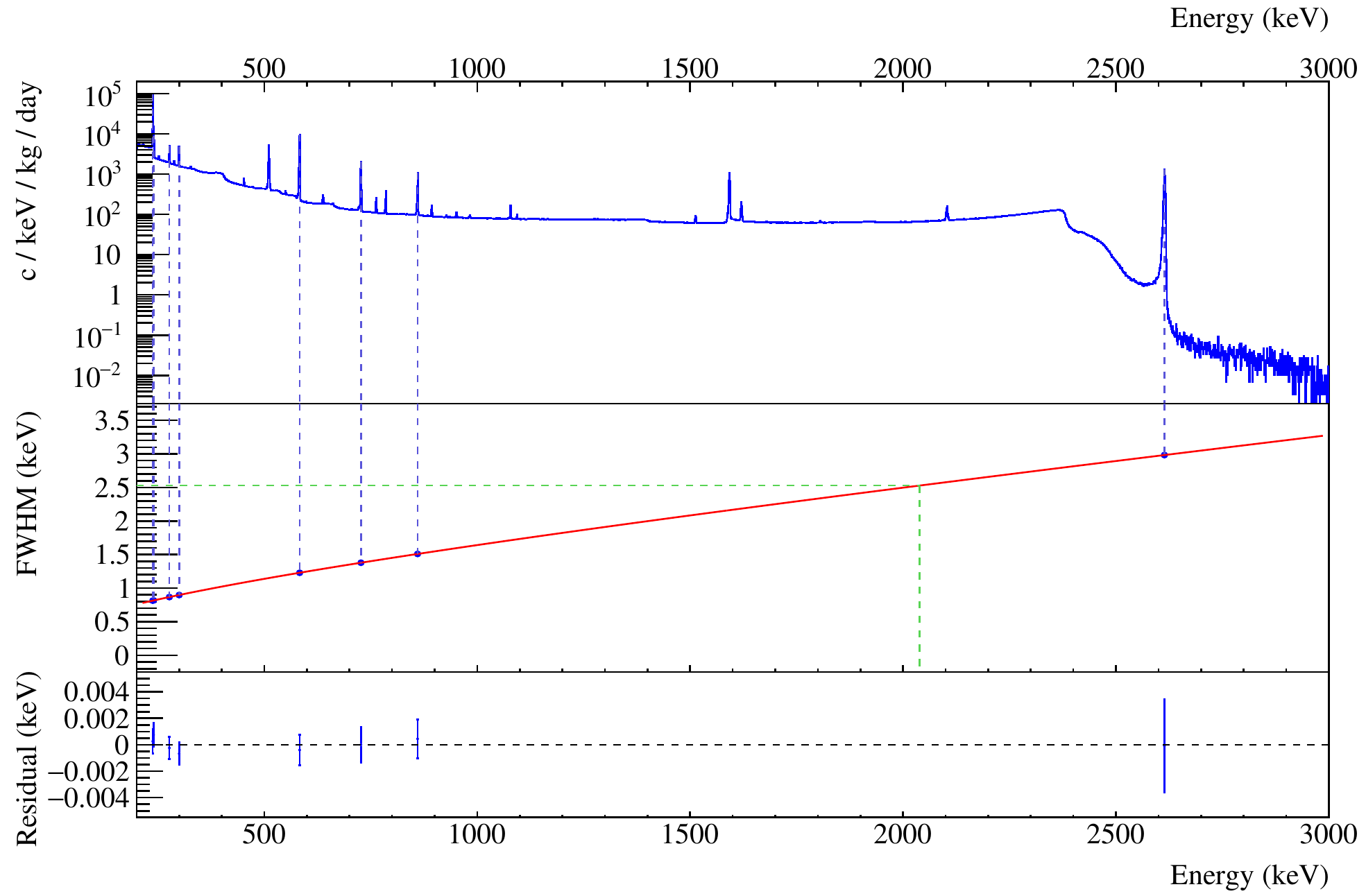}
  \caption{Color online. Top: combined energy spectrum from all DS0-6 calibrations.  Vertical lines indicate the gamma lines used for the final energy calibration in each data set.  Center: exposure-weighted resolution for each gamma peak used in the calibration and a fit to the exposure-weighted values.  The horizontal green line indicates the exposure-weighted average resolution of 2.53~keV at 2039~keV.  Bottom: residuals from the fit in the center panel.\label{fig:eres}}
\end{figure*}

The combined calibration spectrum from DS0-6 is shown in the top panel of Fig.~\ref{fig:eres}.  Using the fitted peak shape, the FWHM at each energy is determined numerically, and the FWHM as a function of energy, $E$,  is fit to
\begin{equation}
  \label{eq:fwhm}
  \text{FWHM}(E) = \sqrt{\Gamma^2_n+\Gamma^2_FE+\Gamma^2_qE^2}.
\end{equation}
The free parameters $\Gamma_n$, $\Gamma_F$, and $\Gamma_q$ account respectively for electronic noise, the Fano factor~\cite{fano1947}, and incomplete charge collection.  The simultaneous fit is performed independently for each data set, and the optimal \nonubb~ROI is then determined using the peak shape parameters evaluated at \qval~and the measured background~\cite{Agostini2017a}.  The optimal ROI width for each data set is shown in Table~\ref{tbl:ds_bg}.  The center panel of Fig.~\ref{fig:eres} shows the exposure-weighted FWHM over all data sets at each of the gamma lines used in the simultaneous fit, along with a fit to Eqn.~\ref{eq:fwhm}.  The fit residuals are shown in the bottom panel.  The exposure-weighted average FWHM at \qval~is $2.53\pm0.08$~keV.

\section{Multi-Site Event Rejection}
\label{sec:avse}

\begin{figure}
  \centering
  \includegraphics[width=0.9\columnwidth]{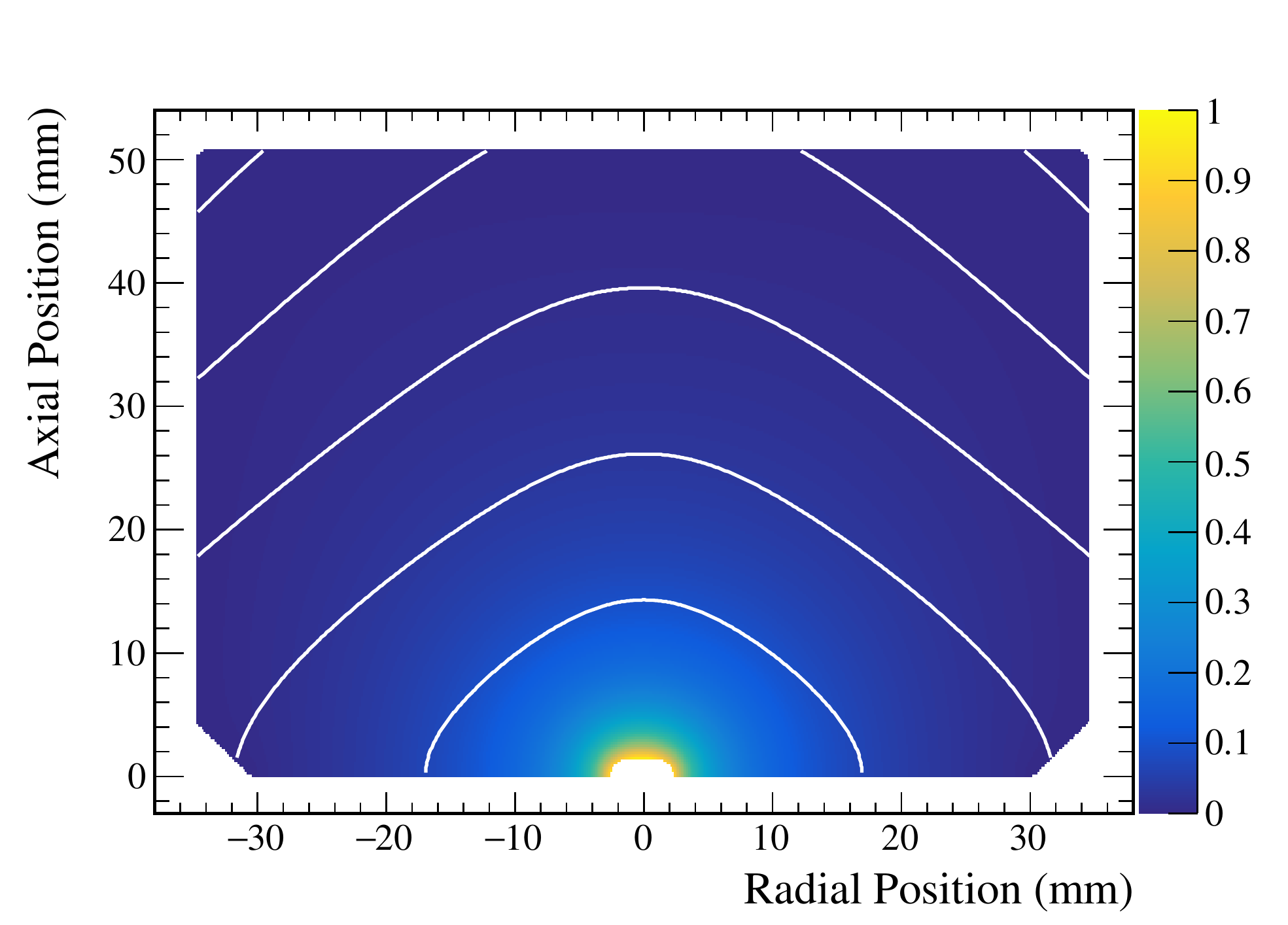}
  \caption{Color online. The value of the weighting potential for a typical enriched PPC detector is indicated by the color scale.  The weighting potential is relatively low in the bulk of the crystal, but quite strong near the point contact at the bottom center.  Lines of equal drift time, separated by 200~ns, are indicated by the white curves.\label{fig:weighting_potential}}
\end{figure}

PPC Ge detectors have a weighting potential~\cite{Shockley,Ramo} that is relatively low in the bulk of the crystal and strongly peaked in the vicinity of the point contact (see Fig.~\ref{fig:weighting_potential}).  The pulse-shape for a bulk interaction has a rise time that is much shorter than the mean drift time, meaning that pulses originating from multiple interaction sites may be resolved.  Due to the limited range of electrons in Ge ($<1$~mm) at the energies of interest, double-beta decays are essentially single-site events.  However, gamma rays of similar energies are likely to interact at multiple sites within a crystal, resulting in pulse-shape differences that can be used to discriminate gamma-ray backgrounds.

\begin{figure}
  \centering
  \includegraphics[width=\columnwidth]{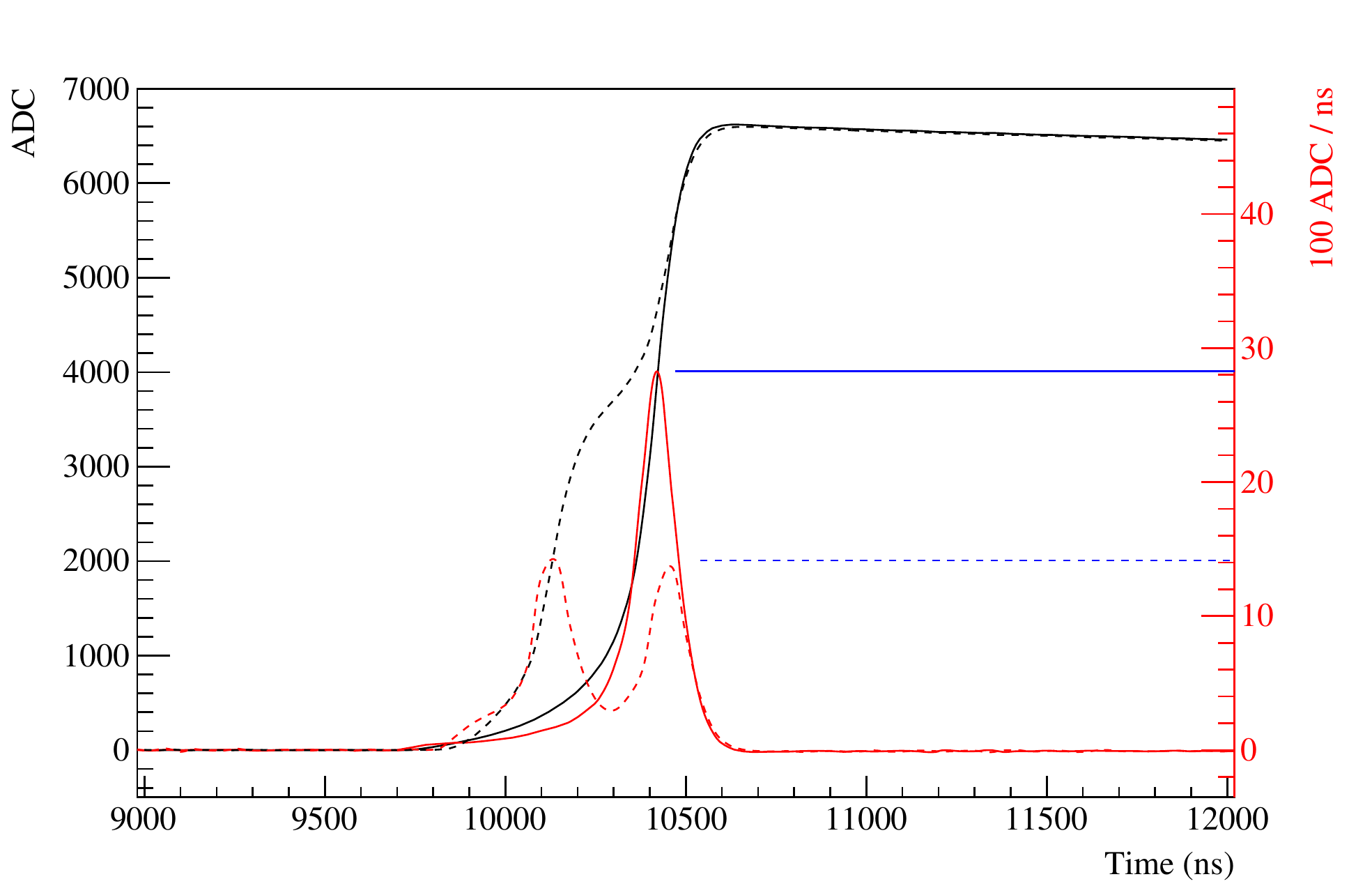}
  \caption{Color online. Shown in black are example single-site (solid) and multi-site (dashed) events from the 2615~keV \nuc{208}{Tl} peak from calibration data for an enriched PPC detector.  The current waveforms are shown in red with blue horizontal lines indicating the maximum current.  While the amplitudes of the voltage waveforms are the same, the maximum current amplitude is significantly lower for multi-site events.\label{fig:avse_wf}}
\end{figure}

\begin{figure}
  \centering
  \includegraphics[width=\columnwidth]{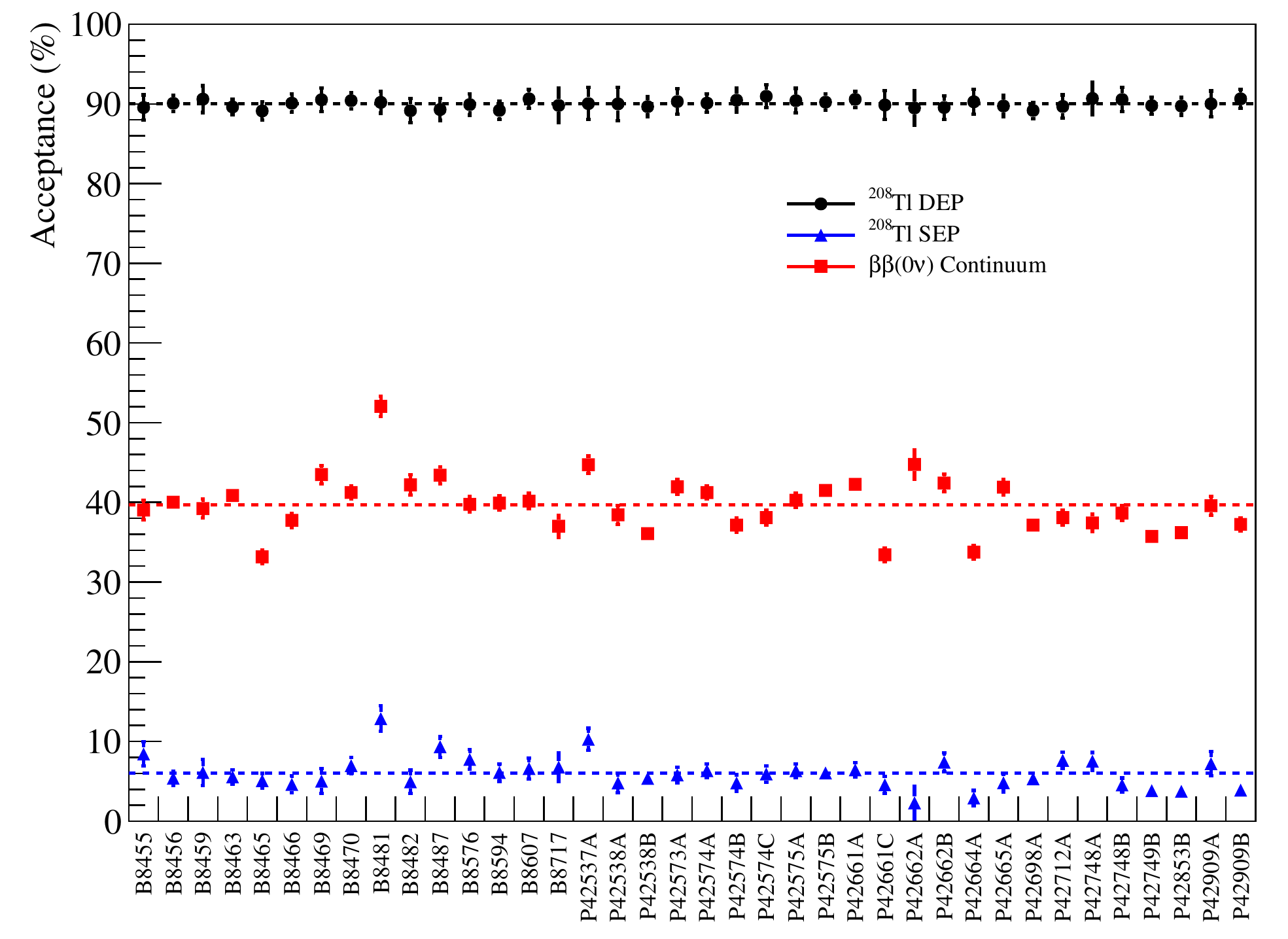}
  \caption{Color online. The acceptance for each detector in DS6a calibration data for events from the 1593~keV DEP and 2104~keV SEP of the 2615~keV \nuc{208}{Tl} decay are shown in black and blue respectively.  Shown in red is the acceptance of Compton scattering events from the calibration source with energy in a 100-keV-wide window centered on the Q-value of 2039~keV.  The errors shown are statistical only, and the horizontal lines indicate the mean value for all calibrated detectors, including natural detectors.  The detector serial numbers are shown as the horizontal axis labels with natural detectors grouped on the left and enriched detectors on the right (serial numbers beginning with `B' and `P' respectively).  Although the detector B8481 has abnormally high acceptance for events outside the DEP, it is a natural detector which is not included in the results of Section~\ref{sec:results}, except for the purposes of rejecting multiple-detector events.\label{fig:avse_acc}}
\end{figure}

Figure~\ref{fig:avse_wf} shows example single-site and multi-site 2615~keV events from a calibration data set for a PPC detector.  While these events have approximately the same reconstructed energy ($E$), the maximum current amplitude ($A$) of the multi-site event (MSE) is degraded relative to the single-site event (SSE).  Calibration data is used to fit a quadratic polynomial to the mean value of $A$ as a function of $E$ for each data set and detector.  To distinguish SSE from MSE, we define the parameter AvsE as
\begin{equation}
  \label{eq:avse}
  \text{AvsE} = \frac{1}{j}\left(p_0 + p_1E + p_2E^2 - \lambda A\right),
\end{equation}
where $\lambda$ is the calibration constant used to convert the measured energy from ADC to keV and $p_0$, $p_1$, and $p_2$ are the coefficients from the quadratic fit to $A$ as a function of $E$.  With the energy dependence removed, $j$ is adjusted to scale AvsE for each detector such that a cut on the parameter above the value of -1 is 90\% efficient in accepting single-site events from the 1593-keV double-escape peak (DEP) from the 2615-keV \nuc{208}{Tl} gamma-ray.  Since these events are single-site, like \nonubb, their acceptance is used as a proxy for the acceptance of \nonubb~events. Figure~\ref{fig:avse_acc} shows the survival percentage for each detector in DS6a \nuc{208}{Tl} calibration data for events in the DEP, single-escape peak (SEP), and the continuum in a 100-keV-wide region centered on the \nonubb~Q-value.  Approximately 6\% of SEP events, which are predominantly multi-site, are retained while 40\% of Compton continuum events are accepted near the \nonubb~ROI.

The signal acceptance in the \nonubb~ROI and its uncertainty are evaluated for each data set utilizing regular calibrations.  The results are listed in Table~\ref{tbl:efficiency}.  The dominant contribution to the systematic uncertainty arises from the difference in the position distribution of simulated \nonubb~events and interactions from the calibration sources, which are used for determining the efficiency.  This is estimated using simulations from the Geant4~\cite{Agostinelli2003}-based MaGe~\cite{Bos11} framework, in addition to the \texttt{siggen}~\cite{RadfordSiggen} package, which is used to simulate detector signal waveforms.  Sub-dominant systematics account for the time-variation in the SSE acceptance, the difference between physics and calibration data, and the energy dependence of the cut acceptance.  Additional detai on the multi-site rejection and estimation of the uncertainties can be found in~\cite{MJD_AvsE}.

\section{Rejection of $\alpha$ Particle Backgrounds}
\label{sec:dcr}

The enriched PPC detectors used in the \MJD~have lithiated dead layers over the surface of the crystals, with the exception of the passivated surface that spans the face with the point contact.  The dead layers have been measured with collimated $^{133}$Ba source scans to be approximately 1.1~mm thick, resulting in a $90\pm1$\% active volume.  External $\alpha$ particles with a few MeV of energy have a range of tens of $\mu$m in Ge.  Therefore, $\alpha$ particles with  the energy of typical decays incident on the lithiated surface cannot penetrate the dead layer of the detector and are not a source of background.  However, $\alpha$ particles impinging on the passivated surface, nominally $\sim$0.1~$\mu$m thick, can deposit energy in an active region of the detector.

Events due to $\alpha$ particles that penetrate the passivated surface  will typically have only a small fraction of the total collected charge due to electrons since the weighting potential is small along the majority of the surface.  The holes in the immediate vicinity of the passivated surface are strongly trapped and subsequently released on timescales much longer than the rise time of events in the bulk, degrading the measured energy.  These energy-degraded, passivated surface $\alpha$ events can then be a potential background near \qval.  However, the slow collection of the holes can be used to discriminate such events from interactions in the crystal bulk~\cite{GruszkoDCR}.

\begin{figure}
  \centering
  \includegraphics[width=\columnwidth]{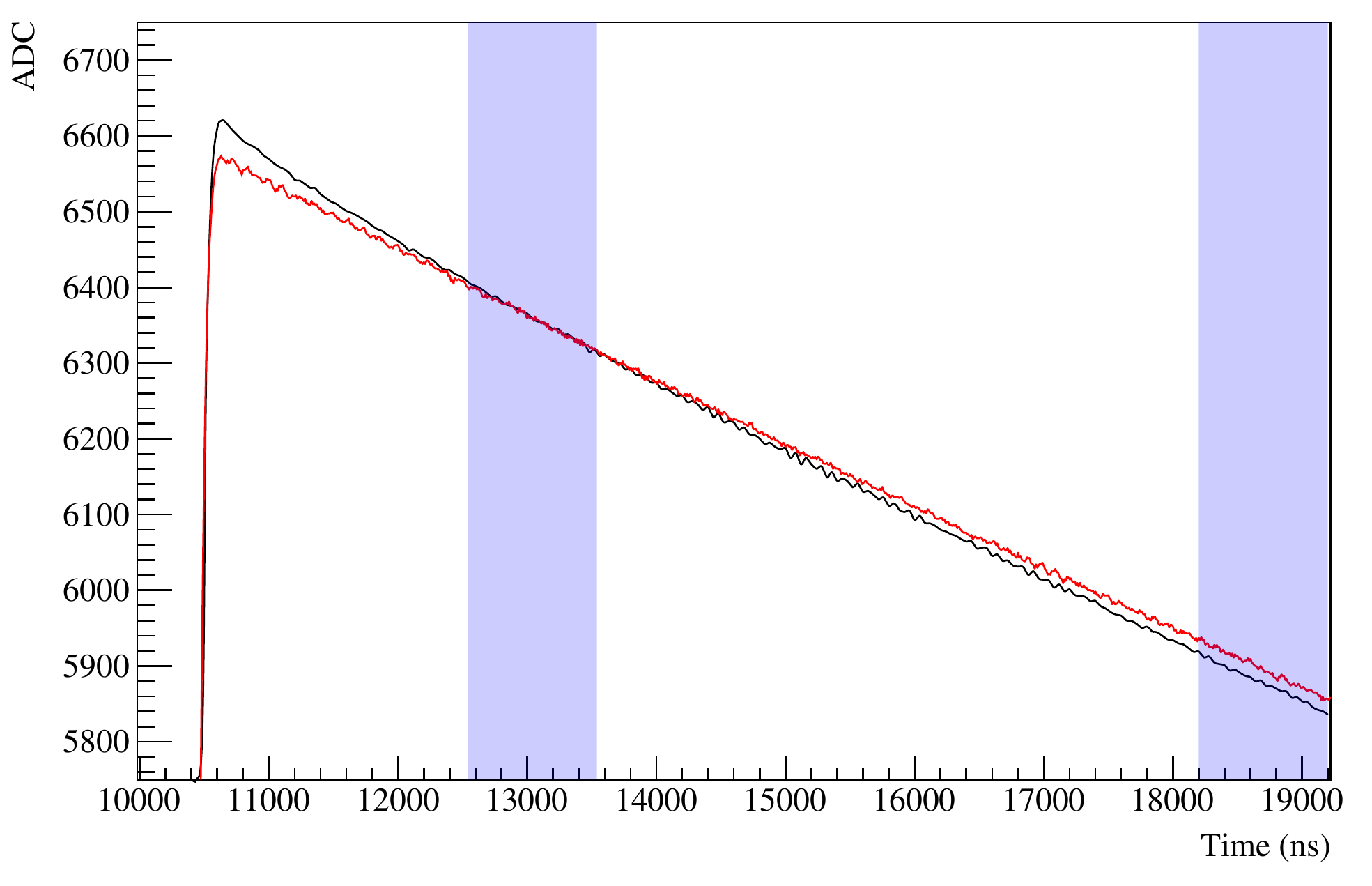}
  \caption{Color online. The single-site waveform from Fig.~\ref{fig:avse_wf} (black) compared to an event in the same detector of the same calibrated energy containing a delayed charge component (red).  The waveforms are aligned at 97\% of the maximum which is the time reference for the shaded regions that are used in computing the DCR slope parameter.\label{fig:dcr_wf}}
\end{figure}

Figure~\ref{fig:dcr_wf} shows the single-site bulk event shown in Fig.~\ref{fig:avse_wf} compared to an energy-degraded $\alpha$ interaction on the passivated surface of approximately the same estimated energy.  The slope between the average value of 1~$\mu$s wide regions beginning 2~$\mu$s after the time the waveform reaches 97\% of the maximum ($t_{97}$) and 1~$\mu$s before the end of the waveform ($t_{max}$), as indicated by the shaded regions in Fig.~\ref{fig:dcr_wf}, is used to compute a simple discriminant.  A cut is defined based on the value of this slope ($\Delta$) that accepts 99.9\% of Compton continuum events near \qval~in calibration data.  For each detector, the `delayed charge recovery' (DCR) parameter is then defined by shifting the raw value of the slope at the cut value to 0:
\begin{equation}
  \label{eq:dcr}
  \text{DCR} = \frac{\int_{t_{max}-1\mu\text{s}}^{t_{max}}V(t)\text{d}t - \int_{t_{97}+2\mu\text{s}}^{t_{97}+3\mu\text{s}}V(t)\text{d}t}{t_{max}-t_{97}} - \Delta,
\end{equation}
where $V(t)$ is the digitized waveform. Similar to AvsE, the acceptance in the \nonubb~ROI and its uncertainty are evaluated for each data set (shown in Table~\ref{tbl:efficiency}).  The systematic uncertainty includes the detector-averaged variation in the parameter between periodic calibrations which is of order 0.1\%.  The asymmetry in the systematic uncertainty arises from a bias towards higher acceptance of SSE compared to Compton continuum events near the same energy, which is estimated using the \nuc{208}{Tl} DEP compared to side-bands near the peak.

\section{Results}
\label{sec:results}

Table~\ref{tbl:efficiency} summarizes the key parameters for each data set described in previous sections as well as the efficiency for containing the full energy of a \nonubb~event within the active volume of the detector ($\epsilon_{cont}$).  Decays occurring close to the crystal surface can deposit some energy within the dead layer, resulting in degradation of the collected charge.  Additionally, bremsstrahlung emission can result in the escape of energy from the active volume of the detector.  Based on MaGe simulations, these effects combined result in a \nonubb~containment efficiency of $0.91\pm0.01$.  The uncertainty accounts for uncertainties in the detector geometry and the difference between simulation and literature values for bremsstrahlung rates and electron range.

\begin{table*}[!htpb]
  \caption{A summary of the key parameters of each data set. The exposure calculation is done independently for each detector. Symmetric uncertainties for the last digits are given in parentheses.  The value of $\epsilon_{res}$ varies slightly for each data set, given the measured peak shape and optimal ROI.  The exposure weighted value over all data sets is $\epsilon_{res}=0.900\pm0.007$.  \label{tbl:efficiency}}
  \begin{center}
    \begin{tabular}{lclccccccc}
      \hline
      Data & Start & Data Set & Active Enr. & Exposure & $\epsilon_{AE}$ & $\epsilon_{DCR}$ & $\epsilon_{cont}$ & $\epsilon_{tot}$ & $N T \epsilon_{tot}\epsilon_{res}$ \\
      Set & Date & Distinction & Mass (kg) & (kg-yr) & & & & & ($10^{24}$ atom yr)\\
      \hline\hline
      DS0 & 6/26/15 & No Inner Cu Shield & 10.69(16) & \phantom01.26(02) & 0.901$_{-0.035}^{+0.032}$ & 0.989$_{-0.002}^{+0.009}$ & 0.908(11) & 0.808$_{-0.033}^{+0.031}$ & \phantom0$6.34_{-0.27}^{+0.25}$ \\
      DS1 & 12/31/15 & Inner Cu Shield added & 11.90(17) & \phantom02.32(04) & 0.901$_{-0.040}^{+0.036}$ & 0.991$_{-0.005}^{+0.010}$&0.909(11) & 0.811$_{-0.038}^{+0.035}$ & \phantom0$11.82_{-0.58}^{+0.53}$ \\
      DS2 & 5/24/16 & Pre-summing & 11.31(16) & \phantom01.22(02) & 0.903$_{-0.037}^{+0.035}$ & 0.986$_{-0.005}^{+0.011}$&0.909(11) & 0.809$_{-0.035}^{+0.034}$ & \phantom0$6.24_{-0.29}^{+0.28}$ \\
      DS3 & 8/25/16 & M1 and M2 installed & 12.63(19) & \phantom01.01(01) & 0.900$_{-0.031}^{+0.030}$ & 0.990$_{-0.003}^{+0.010}$ & 0.909(11) & 0.809$_{-0.030}^{+0.030}$ & \phantom0$5.18_{-0.20}^{+0.20}$ \\
      DS4 & 8/25/16 & M1 and M2 installed & \phantom05.47(08) & \phantom00.28(00) & 0.900$_{-0.034}^{+0.031}$ & 0.992$_{-0.002}^{+0.011}$ & 0.908(10) & 0.809$_{-0.032}^{+0.030}$ & \phantom0$1.47_{-0.06}^{+0.06}$ \\
      DS5a & 10/13/16 & Integrated DAQ (noise) & 17.48(25) & \phantom03.45(05) & 0.900$_{-0.036}^{+0.034}$ & 0.969$_{-0.013}^{+0.013}$ & 0.909(13) & 0.792$_{-0.035}^{+0.034}$ & $17.17_{-0.79}^{+0.76}$ \\
      DS5b & 1/27/17 & Optimized Grounding & 18.44(26) & \phantom01.85(03) & 0.900$_{-0.033}^{+0.031}$ & 0.985$_{-0.005}^{+0.014}$&0.909(13) & 0.805$_{-0.032}^{+0.032}$ & \phantom0$9.46_{-0.39}^{+0.39}$ \\
      DS5c & 3/17/17 & Blind &  18.44(26) & 1.97(03) & 0.900$_{-0.033}^{+0.031}$ & $0.985_{-0.003}^{+0.012}$ & 0.908(11) & 0.806$_{-0.031}^{+0.031}$ &  10.31$_{-0.47}^{+0.47}$ \\
      DS6a & 5/11/17 & Pre-summing, blind &  18.44(26) & 12.67(19) & 0.901$_{-0.032}^{+0.032}$ & $0.990_{-0.002}^{+0.008}$ & 0.908(11) & 0.811$_{-0.030}^{+0.030}$ & 65.10$_{-2.92}^{+2.92}$ \\
      \hline\hline
      Total & (DS0-6)        & & & 26.02(53) & & & & & 133.1$\pm$6.3 \\
      Total & (DS1-4,5b-6) & & & 21.31(41) & & & & & 110.0$\pm$5.1 \\
      \hline
    \end{tabular}
  \end{center}
\end{table*}

The efficiencies of the AvsE cut ($\epsilon_{AE}$) and DCR cuts ($\epsilon_{DCR}$) are combined with $\epsilon_{cont}$ to give the total signal efficiency ($\epsilon_{tot}$) in the second from last column of Table~\ref{tbl:efficiency}.  The total efficiency weighted by exposure for DS0-6 is $0.810^{+0.031}_{-0.032}$.  As described in Section~\ref{sec:ecal}, the ROI for each data set is optimized based on the measured peak shape parameters and background index.  $\epsilon_{res}$ is the fraction of \nonubb~events falling in the optimal ROI (see Table~\ref{tbl:ds_bg}) for a simple counting measurement.  The product of the number of $^{76}$Ge atoms ($N$), the live time ($T$), the total signal efficiency, and $\epsilon_{res}$ is given in the final column of the table.  Taking the exposure weighted mean over all data sets, the \nonubb~ROI containment efficiency is $\epsilon_{res}=0.900\pm0.007$.

As described in Section~\ref{sec:analysis}, some data sets were acquired with fully open data due to construction and commissioning activities.  In total, 11.85~kg-yr of the total 26~kg-yr exposure presented here was blinded across the entire spectrum.  A staged unblinding procedure began on 16 May 2018, with the opening of all data outside of the 1950-2350~keV window used for background estimation near \qval.  The final opening of the $\pm5$~keV window centered on \qval~was completed on 30 May 2018.

\begin{figure}[!htbp]
  \centering
  \includegraphics[width=\columnwidth]{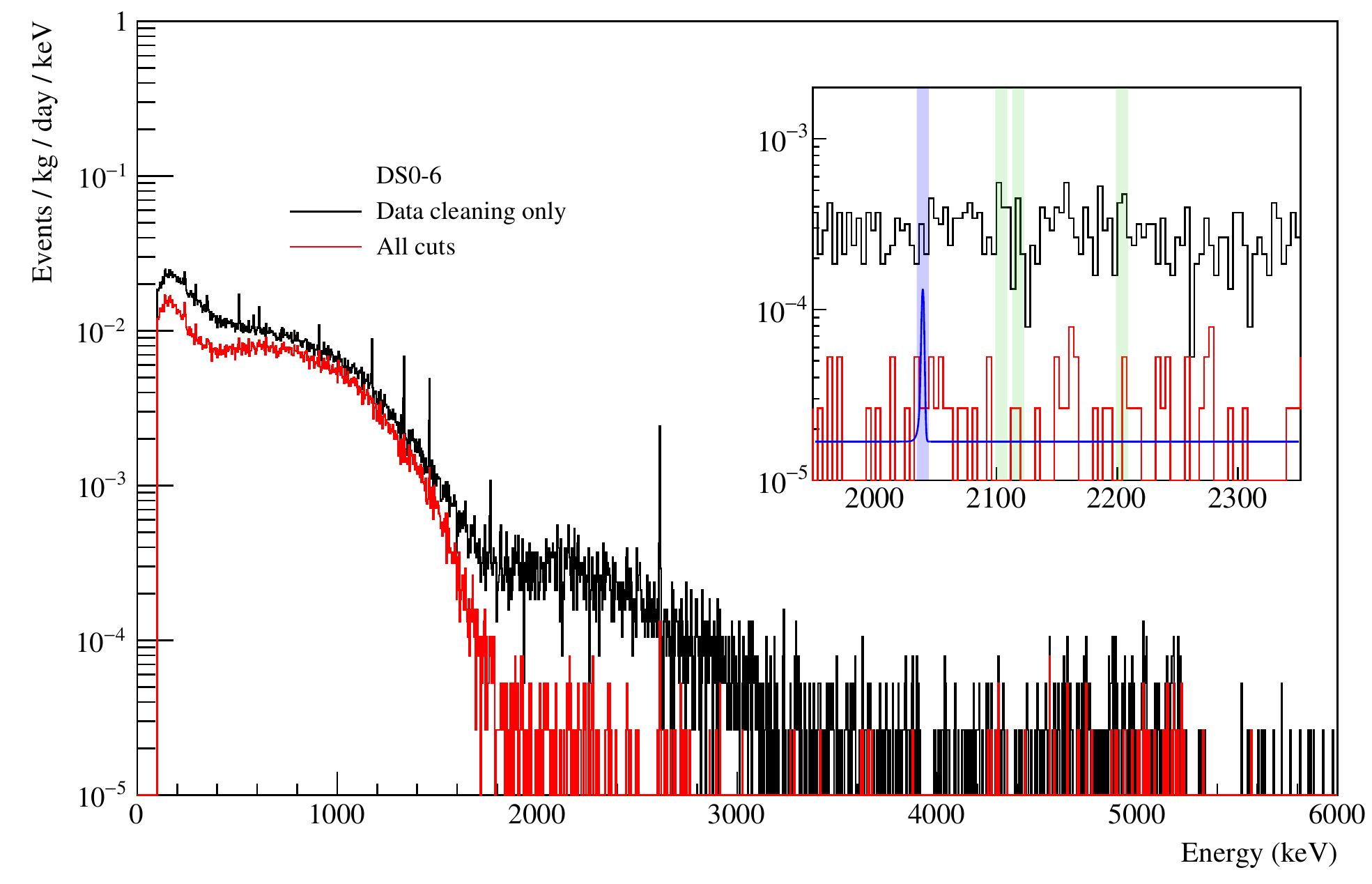}
  \caption{Color online. Energy spectrum above 100~keV of all seven data sets summed together with only data cleaning and muon veto cuts (black) and after all cuts (red).  The inset shows the same spectra in the background estimation window, which spans 1950-2350~keV, with regions excluded due to gamma backgrounds shaded in green and the 10~keV window centered on \qval~shaded in blue.  The solid blue curve shows the flat background estimated from the unshaded regions in the inset plus the 90\% CL upper limit on the number of counts in the \qval~peak for the measured peak shape parameters in each data set weighted by exposure.\label{fig:bg_dc_all}}
\end{figure}

\begin{figure*}[!htbp]
  \centering
  \includegraphics[width=0.85\textwidth]{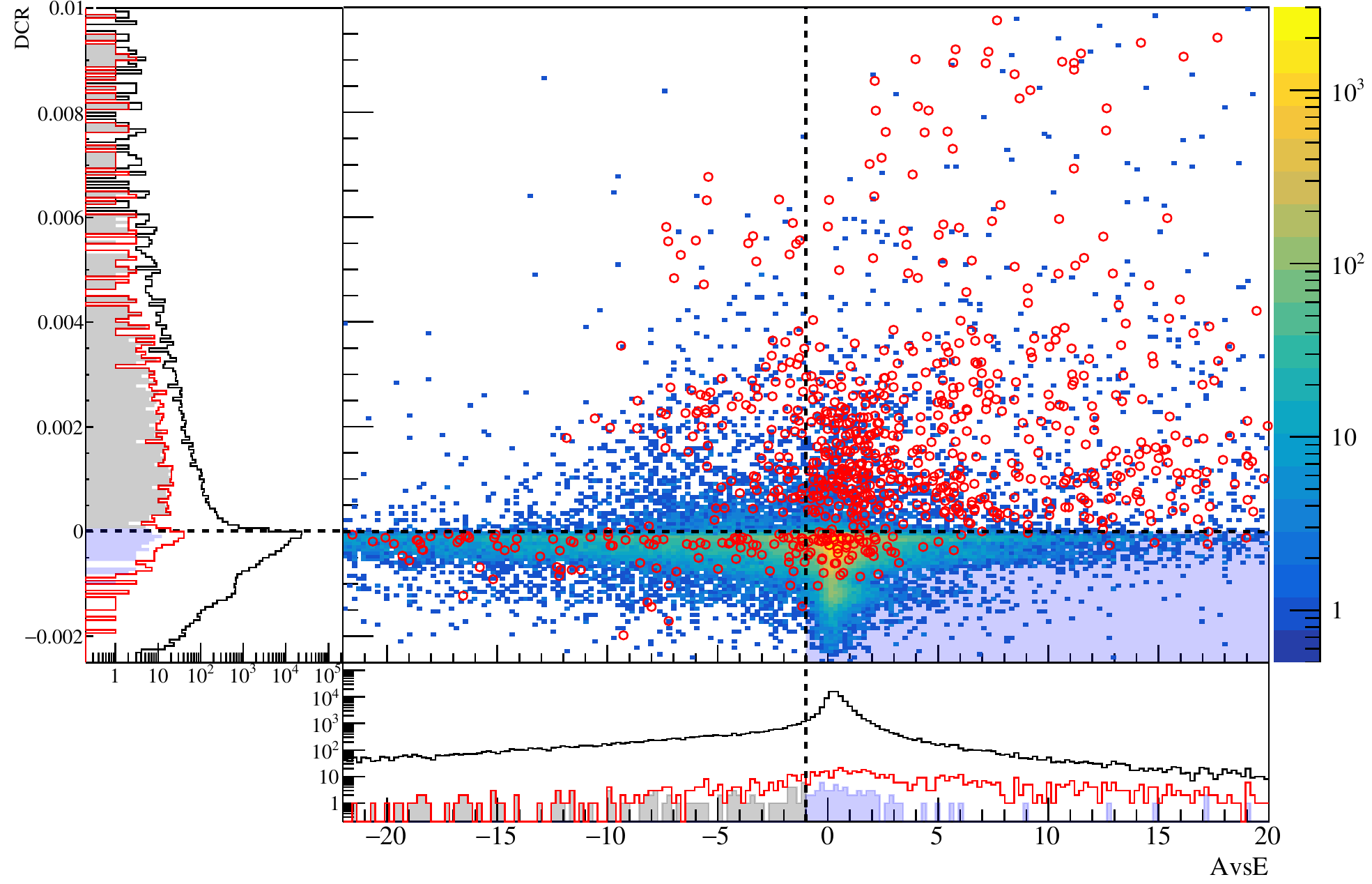}
  \caption{Color online. The distribution of the DCR and AvsE parameters for all background data above 100~keV after data cleaning cuts.  Events with energy between 1950-2350~keV are shown in red.  The calibrated cut values are indicated by the dashed lines.  Events falling in the shaded lower-right quadrant of main panel pass both cuts and are then indicated by blue shading in the projections for events with energy in the range of 1950-2350~keV.  For the same energy range, the events populating the gray shaded regions in the projections pass the cut on the axis projected out but fail the cut on the plotted axis.\label{fig:bg_dcr_avse}}
\end{figure*}

Figure~\ref{fig:bg_dc_all} shows the measured energy spectra above 100~keV for the full enriched detector exposure.  The spectrum shown in black has only data cleaning cuts applied.  The spectrum shown in red also has the coincidence, multi-site, and delayed charge cuts applied, with the latter two responsible for the majority of the difference between the spectra.  Figure~\ref{fig:bg_dcr_avse} shows the DCR and AvsE parameters for all of the background data shown in the data cleaning only spectrum of Fig.~\ref{fig:bg_dc_all}.  Events between 1950-2350~keV (corresponding to the range in the inset of Fig.~\ref{fig:bg_dc_all}) are shown in red.  The cut values are indicated by dashed lines, with the bottom right region containing accepted events.  The DCR cut eliminates the majority of the background in this energy range, and the AvsE cut additionally eliminates multi-site events primarily from \nuc{208}{Tl}.

The inset of Fig.~\ref{fig:bg_dc_all} shows the background spectrum in the energy range from 1950-2350~keV.  MaGe background simulations using assayed component activities predict an approximately flat background in this range with the exception of peaks at 2103~keV, due to the \nuc{208}{Tl} single-escape peak, and at 2118~keV and 2204~keV, due to \nuc{214}{Bi} gamma rays.  For the purposes of background estimation in the ROI, $\pm5$~keV regions centered on these peaks, as indicated by green shading in the inset of Fig.~\ref{fig:bg_dc_all}, are excluded.  Additionally, a $\pm5$~keV wide window centered at \qval~is excluded, as indicated by the blue shaded region in the inset.  After applying all cuts, the background predicted in the ROI from the resulting 360~keV window is $6.1\pm0.8\times10^{-3}$~counts/(keV~kg~yr) or $15.4\pm2.0$~counts/(FWHM~t~yr), using the exposure-weighted optimal ROI of 4.13~keV.  Table~\ref{tbl:ds_bg} summarizes the backgrounds in each data set.

\begin{table}[!htbp]
  \caption{The background (BG) within the 360~keV window defined in the text for each data set.  The background index (BI) is given in units of counts/(keV~kg~yr).  The optimum ROI width for each data set is also given, and the final column shows the resulting expected number of background counts within that ROI.  The second from last row provides a summary for all data sets, and the final row shows the combined total for the lower-background data sets.\label{tbl:ds_bg}}
  \begin{center} 
    \begin{tabular}{lcccc}
      \hline
      Data & Window & BI       & ROI   & ROI BG   \\
      Set  & Counts & $10^{-3}$ & (keV) & (counts) \\
      \hline\hline
      DS0  & 11 & $24.3_{-7.0}^{+8.4}$ & 3.93 & 0.120 \\
      DS1  &  5 &  $6.0_{-2.7}^{+3.4}$ & 4.21 & 0.058 \\
      DS2  &  2 &  $4.6_{-2.9}^{+5.1}$ & 4.34 & 0.024 \\
      DS3  &  0 &              $<$3.6 & 4.39 & 0.000 \\
      DS4  &  0 &             $<$12.7 & 4.25 & 0.000 \\
      DS5a & 10 &  $8.0_{-2.6}^{+3.1}$ & 4.49 & 0.125 \\
      DS5b &  0 &              $<$1.9 & 4.33 & 0.000 \\
      DS5c &  5 &   $7.0_{-3.2}^{+4.0}$ & 4.37 & 0.061 \\
      DS6a & 24 &   $5.3_{-1.0}^{+1.2}$ & 3.93 & 0.262 \\
      \hline\hline
      Total & 57 & $6.1\pm0.8$  & 4.13 & 0.653 \\
      DS1-4,5b-6 & 36 & $4.7\pm0.8$ & 4.14 & 0.529 \\
      \hline
    \end{tabular}
  \end{center}
\end{table}

Also shown in Table~\ref{tbl:ds_bg} is the combined background index from the lower-background configurations, DS1-4,5b-6.  As in~\cite{Aalseth2018}, DS0 is excluded due to the lack of the inner copper shield, and DS5a is excluded due to excess electronic noise.  The background in the lower-background configurations is $11.9\pm2.0$~counts/(FWHM~t~y) or $4.7\pm0.8\times10^{-3}$ counts/(keV~kg~yr)  based on an exposure of 21.3~kg-yr.  Relative to the result with limited statistics presented in~\cite{Aalseth2018} of $4.0_{-2.5}^{+3.1}$~counts/(FWHM~t~yr), this result incorporates a factor of 4 more data and includes blind data selection.  The background near the ROI is largely consistent with \nuc{208}{Tl} contamination in component(s) at larger than assay values.  Investigation into the source of this contamination is ongoing.

The \nonubb~half-life limit set using DS0-6 can be approximated as a Poisson process search in the optimized ROI.  As shown in Table~\ref{tbl:ds_bg}, the expected number of background events in the ROI, given the background index, is 0.65.  The lower-limit on the half-life is given by
\begin{equation}
  \label{eq:thalf}
  T_{1/2}^{0\nu} > \frac{\text{ln}(2)NT\epsilon_{tot}\epsilon_{res}}{S},
\end{equation}
where $S$ is the upper limit on the number of \nonubb~ signal events and $\epsilon_{res}$ is only relevant for the simple counting experiment.  Using the Feldman-Cousins approach~\cite{Feldman1998} with 0.65 expected background events and 1 event observed in the ROI at 2040~keV results in a $^{76}$Ge \nonubb~half-life lower limit of $2.5\times10^{25}$~yr at 90\% CL.

\begin{figure}
  \centering
  \includegraphics[width=\columnwidth]{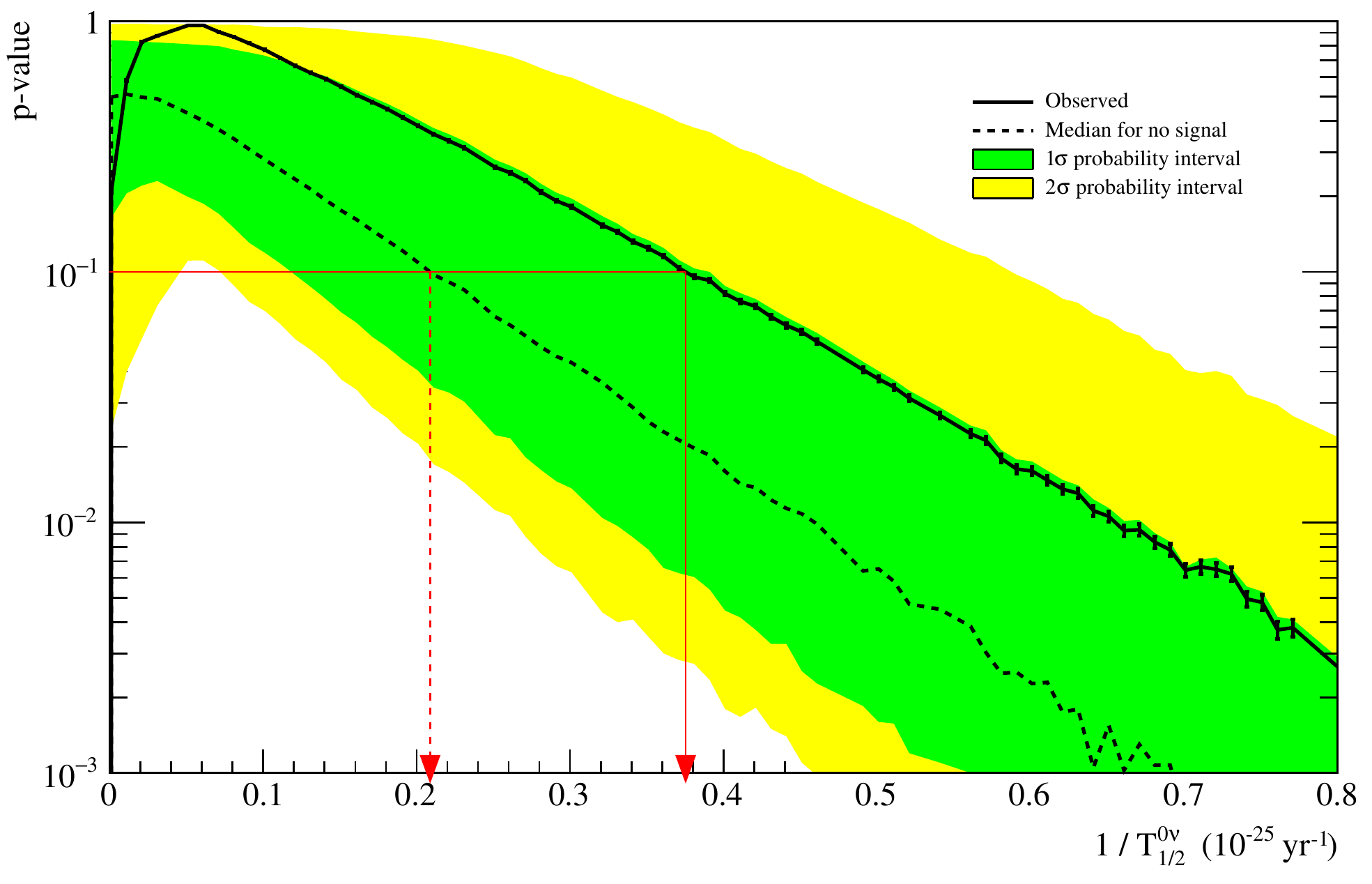}
  \caption{Color online. The p-value as a function of \nonubb~half-life obtained from the unbinned frequentist profile-likelihood method for DS0-6 (solid black).  The \nonubb~half-life in $^{76}$Ge where the p-value of the observed \MJD~data (solid red line) equals 0.1 corresponds to the lower limit on the half-life.  The median sensitivity is indicated with the dashed black line, and the shaded bands correspond to the 1 and 2 $\sigma$ intervals.\label{fig:limit}}
\end{figure}

As in~\cite{Aalseth2018}, we derive our quoted limit using an unbinned, extended profile likelihood method in the RooStats~\cite{Verkerke2003,Schott2012,Agostini2017} framework.  As discussed above and shown in Fig.~\ref{fig:bg_dc_all}, the background is assumed to be flat between 1950-2350~keV with 10~keV-wide regions around potential background peaks removed.  While the supposed \nonubb~half-life is common for all data sets, the peak shape parameters and signal efficiencies are constrained to their data set-specific values as Gaussian nuisance terms.  Monte Carlo simulations were performed for the Neyman interval construction.  Using this method, the median sensitivity at 90\% CL for exclusion is $>4.8\times10^{25}$~yr as shown in Fig.~\ref{fig:limit} with $1\sigma$ and $2\sigma$ contours.  The observed lower limit, based on the measured p-value distribution of the $^{76}$Ge \nonubb~decay half-life, is
\begin{equation*}
  T_{1/2}^{0\nu} > 2.7\times10^{25}\;\text{yr}
\end{equation*}
at 90\% CL, which is also indicated in Fig.~\ref{fig:limit}.  The corresponding upper limit on the number of \nonubb~events at 90\% CL is 3.8, which is shown by the normalization of the blue curve above the flat background in the inset of Fig.~\ref{fig:bg_dc_all}.  The half-life limit is weaker than the median sensitivity by $1\sigma$, largely due to the proximity to \qval~of an observed event at 2040~keV.  As in~\cite{Aalseth2018}, we choose to quote the profile likelihood-based result because it has reliable coverage by construction, based on simulations. GERDA also follows this approach, which facilitates comparison.

As in~\cite{Aalseth2018}, a number of alternative statistical analyses were explored.  A Bayesian statistical analysis was performed using Markov chain Monte Carlo simulations in RooStats with the same likelihood function as above.  With a flat prior on $1/T_{1/2}^{0\nu}$, the Bayesian limit on the half-life is $2.5\times10^{25}$~yr for a 90\% credible interval.  Furthermore, a modified profile likelihood analysis, known as the CL$_{\text{S}}$ method~\cite{Read2000}, yields a \nonubb~half-life lower limit of $2.5\times10^{25}$~yr at 90\% CL.

In order to convert the limit on $T_{1/2}^{0\nu}$ to limits on $\langle m_{\beta\beta}\rangle$, we assume a range of matrix elements in $^{76}$Ge of $2.81<M_{0\nu}<6.13$~\cite{Men09,Horoi2016,Barea2015,Hyvarinen2015,Simkovic2013,Vaquero2013,Yao2015}, phase space factors $(G_{0\nu})$ of $2.36\times10^{-15}$/yr~\cite{Kotila2012} or $2.37\times10^{-15}$/yr~\cite{Mirea2015}, and a value of $g_A=1.27$.  A comprehensive review of the relevant matrix-element theory can be found in~\cite{Engel2016}.  Using these values, our lower limit on $T_{1/2}^{0\nu}$ of $2.7\times10^{25}$ translates into a range of limits on $\langle m_{\beta\beta}\rangle<(200-433)$~meV.

\section{Discussion}
\label{sec:discussion}

The \MJ~Collaboration is currently operating the \DEM, two arrays of HPGe detectors constructed from  ultra-low-background components with the goal of showing that backgrounds can be reduced to a value low enough to justify a tonne-scale \nonubb~experiment using $^{76}$Ge, ultimately with backgrounds at the level of $<0.1$~counts/(FWHM~t~y)~\cite{Abgrall2017d}.  For this result, which includes data up to 16 April 2018, the \DEM~has accrued 26~kg-yr of enriched Ge exposure.   The measured energy resolution at \qval~of $2.53\pm0.08$~keV leads all large-scale \nonubb~experiments to date. The measured background in the low-background configurations (21.3~kg-yr of the total exposure) is $11.9\pm2.0$~counts/(FWHM~t~y). The measured background index, in these units (emphasizing the importance of energy resolution), is second only to the recent GERDA result~\cite{Zsigmond2018} of 1.8 or 2.2~counts/(FWHM~t~y) for Phase II BEGE and coaxial style detectors respectively.

With the full exposure of 26~kg-yr, the \DEM~has reached a limit on the \nonubb~half-life in $^{76}$Ge of $2.7\times10^{25}$~yr at 90\% CL with a median sensitivity of $4.8\times10^{25}$~yr (90\% CL).  The present leading half-life limit for $^{76}$Ge has been reported by GERDA~\cite{Zsigmond2018} at $9\times10^{25}$~yr (90\% CL) from 82.4~kg-yr of exposure.  A combined limit from these two Ge-based experiments would, at present, exceed $10^{26}$~yr.

\section*{Acknowledgments}
The authors appreciate the technical assistance of J.F. Amsbaugh, J. Bell, B.A. Bos, T.H. Burritt, G. Capps, K. Carney, R. Daniels, L. DeBraeckeleer, C. Dunagan, G.C. Harper, C. Havener, G. Holman, R. Hughes, K. Jeskie, K. Lagergren, D. Lee, M. Middlebook, A. Montoya, A.W. Myers, D.Peterson, D. Reid, L. Rodriguez, H. Salazar, A.R. Smith, G. Swift, M. Turqueti, J. Thompson, P. Thompson, C. Tysor, T.D. Van Wechel, R. Witharm, and H. Yaver.

This material is based upon work supported by the U.S.~Department of Energy, Office of Science, Office of Nuclear Physics under Award Numbers DE-AC02-05CH11231,  DE-AC05-00OR22725, DE-AC05-76RL0130, DE-AC52-06NA25396, DE-FG02-97ER41020, DE-FG02-97ER41033, DE-FG02-97ER41041, DE-SC0010254, DE-SC0012612, DE-SC0014445, and DE-SC0018060. We acknowledge support from the Particle Astrophysics Program and Nuclear Physics Program of the National Science Foundation through grant numbers MRI-0923142, PHY-1003399, PHY-1102292, PHY-1206314, PHY-1614611, PHY-1812409, and PHY-1812356. We gratefully acknowledge the support of the U.S.~Department of Energy through the LANL/LDRD Program and through the PNNL/LDRD Program for this work. We acknowledge support from the Russian Foundation for Basic Research, grant No.~15-02-02919. We acknowledge the support of the Natural Sciences and Engineering Research Council of Canada, funding reference number SAPIN-2017-00023, and from the Canada Foundation from Innovation John R.~Evans Leaders Fund.  This research used resources provided by the Oak Ridge Leadership Computing Facility at Oak Ridge National Laboratory and by the National Energy Research Scientific Computing Center, a U.S.~Department of Energy Office of Science User Facility. We thank our hosts and colleagues at the Sanford Underground Research Facility for their support.

\bibliographystyle{iopart-num}
\bibliography{nu_2018_0vbb}

\end{document}

%% file: mjd_authors.tex
\newcommand{\ITEP}{National Research Center ``Kurchatov Institute'' Institute for Theoretical and Experimental Physics, Moscow, Russia}
\newcommand{\JINR}{Joint Institute for Nuclear Research, Dubna, Russia}
\newcommand{\lbnl}{Nuclear Science Division, Lawrence Berkeley National Laboratory, Berkeley, CA, USA}
\newcommand{\lanl}{Los Alamos National Laboratory, Los Alamos, NM, USA}
\newcommand{\queens}{Department of Physics, Engineering Physics and Astronomy, Queen's University, Kingston, ON, Canada}
\newcommand{\uw}{Center for Experimental Nuclear Physics and Astrophysics, 
and Department of Physics, University of Washington, Seattle, WA, USA}
\newcommand{\unc}{Department of Physics and Astronomy, University of North Carolina, Chapel Hill, NC, USA}
\newcommand{\duke}{Department of Physics, Duke University, Durham, NC, USA}
\newcommand{\ncsu}{Department of Physics, North Carolina State University, Raleigh, NC, USA}
\newcommand{\ornl}{Oak Ridge National Laboratory, Oak Ridge, TN, USA}
\newcommand{\ou}{Research Center for Nuclear Physics, Osaka University, Ibaraki, Osaka, Japan}
\newcommand{\pnnl}{Pacific Northwest National Laboratory, Richland, WA, USA}
\newcommand{\princeton}{Department of Physics, Princeton University, Princeton, NJ, USA}
\newcommand{\ttu}{Tennessee Tech University, Cookeville, TN, USA}
\newcommand{\sdsmt}{South Dakota School of Mines and Technology, Rapid City, SD, USA}
\newcommand{\usc}{Department of Physics and Astronomy, University of South Carolina, Columbia, SC, USA}
\newcommand{\usd}{Department of Physics, University of South Dakota, Vermillion, SD, USA}
\newcommand{\ut}{Department of Physics and Astronomy, University of Tennessee, Knoxville, TN, USA}
\newcommand{\tunl}{Triangle Universities Nuclear Laboratory, Durham, NC, USA}
\newcommand{\mpi}{Max-Planck-Institut f\"{u}r Physik, M\"{u}nchen, Germany}
\newcommand{\tum}{Physik Department, Technische Universit\"{a}t, M\"{u}nchen, Germany}
\newcommand{\MIT}{Department of Physics, Massachusetts Institute of Technology, Cambridge, MA, USA}

\affiliation{\uw}
\affiliation{\pnnl}
\affiliation{\usc}
\affiliation{\ornl}
\affiliation{\ITEP}
\affiliation{\usd}
\affiliation{\queens}
\affiliation{\sdsmt}
\affiliation{\duke}
\affiliation{\tunl}
\affiliation{\unc}
\affiliation{\lbnl}
\affiliation{\lanl}
\affiliation{\ut}
\affiliation{\ou}
\affiliation{\princeton}
\affiliation{\ncsu}
\affiliation{\MIT}
\affiliation{\ttu}
\affiliation{\mpi}
\affiliation{\tum}
\affiliation{\JINR}

\author{S.I.~Alvis}\affiliation{\uw}
\author{I.J.~Arnquist}\affiliation{\pnnl}
\author{F.T.~Avignone~III}\affiliation{\usc}\affiliation{\ornl}
\author{A.S.~Barabash}\affiliation{\ITEP}
\author{C.J.~Barton}\affiliation{\usd}
\author{V. Basu}\affiliation{\queens}
\author{F.E.~Bertrand}\affiliation{\ornl}
\author{B.~Bos}\affiliation{\sdsmt}
\author{M.~Busch}\affiliation{\duke}\affiliation{\tunl}
\author{M.~Buuck}\affiliation{\uw}
\author{T.S.~Caldwell}\affiliation{\unc}\affiliation{\tunl}
\author{Y-D.~Chan}\affiliation{\lbnl}
\author{C.D.~Christofferson}\affiliation{\sdsmt}
\author{P.-H.~Chu}\affiliation{\lanl}
\author{C.~Cuesta}\altaffiliation{Present address: Centro de Investigaciones Energ\'{e}ticas, Medioambientales y Tecnol\'{o}gicas, CIEMAT 28040, Madrid, Spain}\affiliation{\uw}
\author{J.A.~Detwiler}\affiliation{\uw}
\author{Yu.~Efremenko}\affiliation{\ut}\affiliation{\ornl}
\author{H.~Ejiri}\affiliation{\ou}
\author{S.R.~Elliott}\affiliation{\lanl}
\author{T.~Gilliss}\affiliation{\unc}\affiliation{\tunl}
\author{G.K.~Giovanetti}\affiliation{\princeton}
\author{M.P.~Green}\affiliation{\ncsu}\affiliation{\tunl}\affiliation{\ornl}
\author{J.~Gruszko}\affiliation{\MIT}
\author{I.S.~Guinn}\affiliation{\uw}
\author{V.E.~Guiseppe}\affiliation{\usc}
\author{C.R.~Haufe}\affiliation{\unc}\affiliation{\tunl}
\author{R.J.~Hegedus}\affiliation{\unc}\affiliation{\tunl}
\author{L.~Hehn}\affiliation{\lbnl}
\author{R.~Henning}\affiliation{\unc}\affiliation{\tunl}
\author{D.~Hervas~Aguilar}\affiliation{\unc}\affiliation{\tunl}
\author{E.W.~Hoppe}\affiliation{\pnnl}
\author{M.A.~Howe}\affiliation{\unc}\affiliation{\tunl}
\author{M.F.~Kidd}\affiliation{\ttu}
\author{S.I.~Konovalov}\affiliation{\ITEP}
\author{R.T.~Kouzes}\affiliation{\pnnl}
\author{A.M.~Lopez}\affiliation{\ut}
\author{R.D.~Martin}\affiliation{\queens}
\author{R.~Massarczyk}\affiliation{\lanl}
\author{S.J.~Meijer}\affiliation{\unc}\affiliation{\tunl}
\author{S.~Mertens}\affiliation{\mpi}\affiliation{\tum}
\author{J.~Myslik}\affiliation{\lbnl}
\author{G.~Othman}\affiliation{\unc}\affiliation{\tunl}
\author{W.~Pettus}\affiliation{\uw}
\author{A.~Piliounis}\affiliation{\queens}
\author{A.W.P.~Poon}\affiliation{\lbnl}
\author{D.C.~Radford}\affiliation{\ornl}
\author{J.~Rager}\affiliation{\unc}\affiliation{\tunl}
\author{A.L.~Reine}\affiliation{\unc}\affiliation{\tunl}
\author{K.~Rielage}\affiliation{\lanl}
\author{N.W.~Ruof}\affiliation{\uw}
\author{B.~Shanks}\affiliation{\ornl}
\author{M.~Shirchenko}\affiliation{\JINR}
\author{D.~Tedeschi}\affiliation{\usc}
\author{R.L.~Varner}\affiliation{\ornl}
\author{S.~Vasilyev}\affiliation{\JINR}
\author{B.R.~White}\affiliation{\lanl}
\author{J.F.~Wilkerson}\affiliation{\unc}\affiliation{\tunl}\affiliation{\ornl}
\author{C.~Wiseman}\affiliation{\uw}
\author{W.~Xu}\affiliation{\usd}
\author{E.~Yakushev}\affiliation{\JINR}
\author{C.-H.~Yu}\affiliation{\ornl}
\author{V.~Yumatov}\affiliation{\ITEP}
\author{I.~Zhitnikov}\affiliation{\JINR}
\author{B.X.~Zhu}\affiliation{\lanl}
			
\collaboration{{\sc{Majorana}} Collaboration}
\noaffiliation